\documentstyle[twoside]{article}
\input{psfig}
\catcode`\@=11
\long\def\@makefntext#1{
\protect\noindent \hbox to 3.2pt {\hskip-.9pt  
$^{{\eightrm\@thefnmark}}$\hfil}#1\hfill}		

\def\@makefnmark{\hbox to 0pt{$^{\@thefnmark}$\hss}}	
	
\def\ps@myheadings{\let\@mkboth\@gobbletwo
\def\@oddhead{\hbox{}
\rightmark\hfil\eightrm\thepage}   
\def\@oddfoot{}\def\@evenhead{\eightrm\thepage\hfil
\leftmark\hbox{}}\def\@evenfoot{}
\def\sectionmark##1{}\def\subsectionmark##1{}}

\oddsidemargin=\evensidemargin
\newcounter{sectionc}\newcounter{subsectionc}\newcounter{subsubsectionc}
\renewcommand{\section}[1] {\vspace{12pt}\addtocounter{sectionc}{1} 
\setcounter{subsectionc}{0}\setcounter{subsubsectionc}{0}\noindent 
	{\tenbf\thesectionc. #1}\par\vspace{5pt}}
\renewcommand{\subsection}[1] {\vspace{12pt}\addtocounter{subsectionc}{1} 
	\setcounter{subsubsectionc}{0}\noindent 
	{\bf\thesectionc.\thesubsectionc. {\kern1pt \bfit #1}}\par\vspace{5pt}}
\renewcommand{\subsubsection}[1] {\vspace{12pt}\addtocounter{subsubsectionc}{1}
	\noindent{\tenrm\thesectionc.\thesubsectionc.\thesubsubsectionc.
	{\kern1pt \tenit #1}}\par\vspace{5pt}}
\newcommand{\nonumsection}[1] {\vspace{12pt}\noindent{\tenbf #1}
	\par\vspace{5pt}}

\newcounter{appendixc}
\newcounter{subappendixc}[appendixc]
\newcounter{subsubappendixc}[subappendixc]
\renewcommand{\thesubappendixc}{\Alph{appendixc}.\arabic{subappendixc}}
\renewcommand{\thesubsubappendixc}
	{\Alph{appendixc}.\arabic{subappendixc}.\arabic{subsubappendixc}}

\renewcommand{\appendix}[1] {\vspace{12pt}
        \refstepcounter{appendixc}
        \setcounter{figure}{0}
        \setcounter{table}{0}
        \setcounter{lemma}{0}
        \setcounter{theorem}{0}
        \setcounter{corollary}{0}
        \setcounter{definition}{0}
        \setcounter{equation}{0}
        \renewcommand{\thefigure}{\Alph{appendixc}.\arabic{figure}}
        \renewcommand{\thetable}{\Alph{appendixc}.\arabic{table}}
        \renewcommand{\theappendixc}{\Alph{appendixc}}
        \renewcommand{\thelemma}{\Alph{appendixc}.\arabic{lemma}}
        \renewcommand{\thetheorem}{\Alph{appendixc}.\arabic{theorem}}
        \renewcommand{\thedefinition}{\Alph{appendixc}.\arabic{definition}}
        \renewcommand{\thecorollary}{\Alph{appendixc}.\arabic{corollary}}
        \renewcommand{\theequation}{\Alph{appendixc}.\arabic{equation}}
        \noindent{\tenbf Appendix \theappendixc #1}\par\vspace{5pt}}
\newcommand{\subappendix}[1] {\vspace{12pt}
        \refstepcounter{subappendixc}
        \noindent{\bf Appendix \thesubappendixc. {\kern1pt \bfit #1}}
	\par\vspace{5pt}}
\newcommand{\subsubappendix}[1] {\vspace{12pt}
        \refstepcounter{subsubappendixc}
        \noindent{\rm Appendix \thesubsubappendixc. {\kern1pt \tenit #1}}
	\par\vspace{5pt}}

\newcommand{\textlineskip}{\baselineskip=13pt}
\newcommand{\smalllineskip}{\baselineskip=10pt}

\def\eightcirc{
\begin{picture}(0,0)
\put(4.4,1.8){\circle{6.5}}
\end{picture}}
\def\eightcopyright{\eightcirc\kern2.7pt\hbox{\eightrm c}} 

\newcommand{\copyrightheading}[1]
	{\vspace*{-2.5cm}\smalllineskip{\flushleft
	{\footnotesize International Journal of Modern Physics A, #1}\\
	{\footnotesize $\eightcopyright$\, World Scientific Publishing
	 Company}\\
	 }}

\def\abstracts#1#2#3{{
	\centering{\begin{minipage}{4.5in}\baselineskip=10pt\footnotesize
	\parindent=0pt #1\par 
	\parindent=15pt #2\par
	\parindent=15pt #3
	\end{minipage}}\par}} 

\renewenvironment{thebibliography}[1]
	{\frenchspacing
	 \ninerm\baselineskip=11pt
	 \begin{list}{\arabic{enumi}.}
	{\usecounter{enumi}\setlength{\parsep}{0pt}
	 \setlength{\leftmargin 12.7pt}{\rightmargin 0pt} 
	 \setlength{\itemsep}{0pt} \settowidth
	{\labelwidth}{#1.}\sloppy}}{\end{list}}

\newcounter{itemlistc}
\newcounter{romanlistc}
\newcounter{alphlistc}
\newcounter{arabiclistc}

\newcommand{\fcaption}[1]{
        \refstepcounter{figure}
        \setbox\@tempboxa = \hbox{\footnotesize Fig.~\thefigure. #1}
        \ifdim \wd\@tempboxa > 5in
           {\begin{center}
        \parbox{5in}{\footnotesize\smalllineskip Fig.~\thefigure. #1}
            \end{center}}
        \else
             {\begin{center}
             {\footnotesize Fig.~\thefigure. #1}
              \end{center}}
        \fi}

\newcommand{\tcaption}[1]{
        \refstepcounter{table}
        \setbox\@tempboxa = \hbox{\footnotesize Table~\thetable. #1}
        \ifdim \wd\@tempboxa > 5in
           {\begin{center}
        \parbox{5in}{\footnotesize\smalllineskip Table~\thetable. #1}
            \end{center}}
        \else
             {\begin{center}
             {\footnotesize Table~\thetable. #1}
              \end{center}}
        \fi}

\def\@citex[#1]#2{\if@filesw\immediate\write\@auxout
	{\string\citation{#2}}\fi
\def\@citea{}\@cite{\@for\@citeb:=#2\do
	{\@citea\def\@citea{,}\@ifundefined
	{b@\@citeb}{{\bf ?}\@warning
	{Citation `\@citeb' on page \thepage \space undefined}}
	{\csname b@\@citeb\endcsname}}}{#1}}

\newif\if@cghi
\def\cite{\@cghitrue\@ifnextchar [{\@tempswatrue
	\@citex}{\@tempswafalse\@citex[]}}
\def\citelow{\@cghifalse\@ifnextchar [{\@tempswatrue
	\@citex}{\@tempswafalse\@citex[]}}
\def\@cite#1#2{{$\null^{#1}$\if@tempswa\typeout
	{IJCGA warning: optional citation argument 
	ignored: `#2'} \fi}}

\def\pmb#1{\setbox0=\hbox{#1}
	\kern-.025em\copy0\kern-\wd0
	\kern.05em\copy0\kern-\wd0
	\kern-.025em\raise.0433em\box0}

\def\fnt#1#2{\footnotetext{\kern-.3em
	{$^{\mbox{\scriptsize #1}}$}{#2}}}

\def\fpage#1{\begingroup
\voffset=.3in
\thispagestyle{empty}\begin{table}[b]\centerline{\footnotesize #1}
	\end{table}\endgroup}

\def\runninghead#1#2{\pagestyle{myheadings}
\markboth{{\protect\footnotesize\it{\quad #1}}\hfill}
{\hfill{\protect\footnotesize\it{#2\quad}}}}
\font\tenrm=cmr10
\font\tenit=cmti10 
\font\tenbf=cmbx10
\font\bfit=cmbxti10 at 10pt
\font\ninerm=cmr9

\font\eightrm=cmr8

\def\qed{\hbox{${\vcenter{\vbox{			
   \hrule height 0.4pt\hbox{\vrule width 0.4pt height 6pt
   \kern5pt\vrule width 0.4pt}\hrule height 0.4pt}}}$}}

\begin{document}

\runninghead{Instructions for Typesetting Camera-Ready
Manuscripts $\ldots$} {Instructions for Typesetting Camera-Ready
Manuscripts $\ldots$}

\normalsize\textlineskip
\thispagestyle{empty}
\setcounter{page}{1}

\copyrightheading{}			

\vspace*{0.88truein}

\fpage{1}
\centerline{\bf  HIGHER-FOCK STATES AND THE COVARIANCE}
\vspace*{0.035truein}
\centerline{\bf  OF THE LIGHT-FRONT QUARK MODEL}
\vspace*{0.37truein}
\centerline{\footnotesize HO-MEOYNG CHOI\footnote{
Present address: Physics Department, Carnegie Mellon University,
Pittsburgh, Pennsylvania 15213, USA.}}
\vspace*{0.015truein}
\centerline{\footnotesize\it Department of Physics, North Carolina
State University}
\baselineskip=10pt
\centerline{\footnotesize\it Raleigh, North Carolina 27695-8202,
USA}
\vspace*{10pt}
\centerline{\footnotesize CHUENG-RYONG JI}
\vspace*{0.015truein}
\centerline{\footnotesize\it Department of Physics, North Carolina
State University}
\baselineskip=10pt
\centerline{\footnotesize\it  Raleigh, North Carolina 27695-8202, USA}

\vspace*{0.21truein}
\abstracts{
The light-front quark model has been successful in describing various 
meson properties. We apply the model to the meson weak decays and discuss 
the covariance of the model. We find that the inclusion of the higher-Fock 
state(i.e. nonvalence or so called Z-graph) contributions is necessary to 
recover the covariance of the light-front quark model.
We present the possibility of the effective calculation of the higher-Fock 
state contributions to restore the covarince in the light-front model. }{}{}

\textlineskip			
\vspace*{12pt}			

\noindent
Perhaps, one of the popular formulations for the analysis of
such exclusive processes may be provided in the framework of light-front
(LF) quantization. In particular, the Drell-Yan-West ($q^+=q^0+q^3=0$) 
frame has been extensively used in the calculation of various electroweak 
form factors and decay processes. As an example,
only the parton-number-conserving (valence) Fock state contribution
is needed in $q^+=0$ frame when the ``good" component of the current,
$J^+$ or ${\bf J}_{\perp}=(J_x,J_y)$, is used for the spacelike
electromagnetic form factor calculation of pseudoscalar mesons.
The LF approach may also provide a bridge between the two
fundamentally different pictures of hadronic matter, i.e. the
constituent quark model (CQM) (or the quark parton model) closely
related to the experimental observations and the quantum chromodynamics
(QCD) based on a covariant non-abelian quantum field theory.
The crux of possible connection between the two pictures is the rational 
energy-momentum dispersion relation that leads to a relatively simple
vacuum structure. There is no spontaneous creation of massive fermions
in the LF quantized vacuum. Thus, one can immediately obtain a
constituent-type picture, in which all partons in a hadronic state are
connected directly to the hadron instead of being simply disconnected
excitations (or vacuum fluctuations) in a complicated medium.

On the other hand, the analysis of timelike exclusive processes (or
timelike $q^2>0$ region of bound-state form factors) remained
as a rather significant challenge in the LF approach. In principle, the
$q^+\neq0$ frame can be used to compute the timelike processes but
then it is inevitable to encounter the particle-number-nonconserving
Fock state (or nonvalence) contribution.
The main source of difficulty in CQM phenomenology
is the lack of information on the black blob in the nonvalence
diagram arising from the quark-antiquark pair creation (see Fig.~1(a)).

\begin{figure}[htbp]
\vspace*{13pt}
\centerline{\psfig{figure=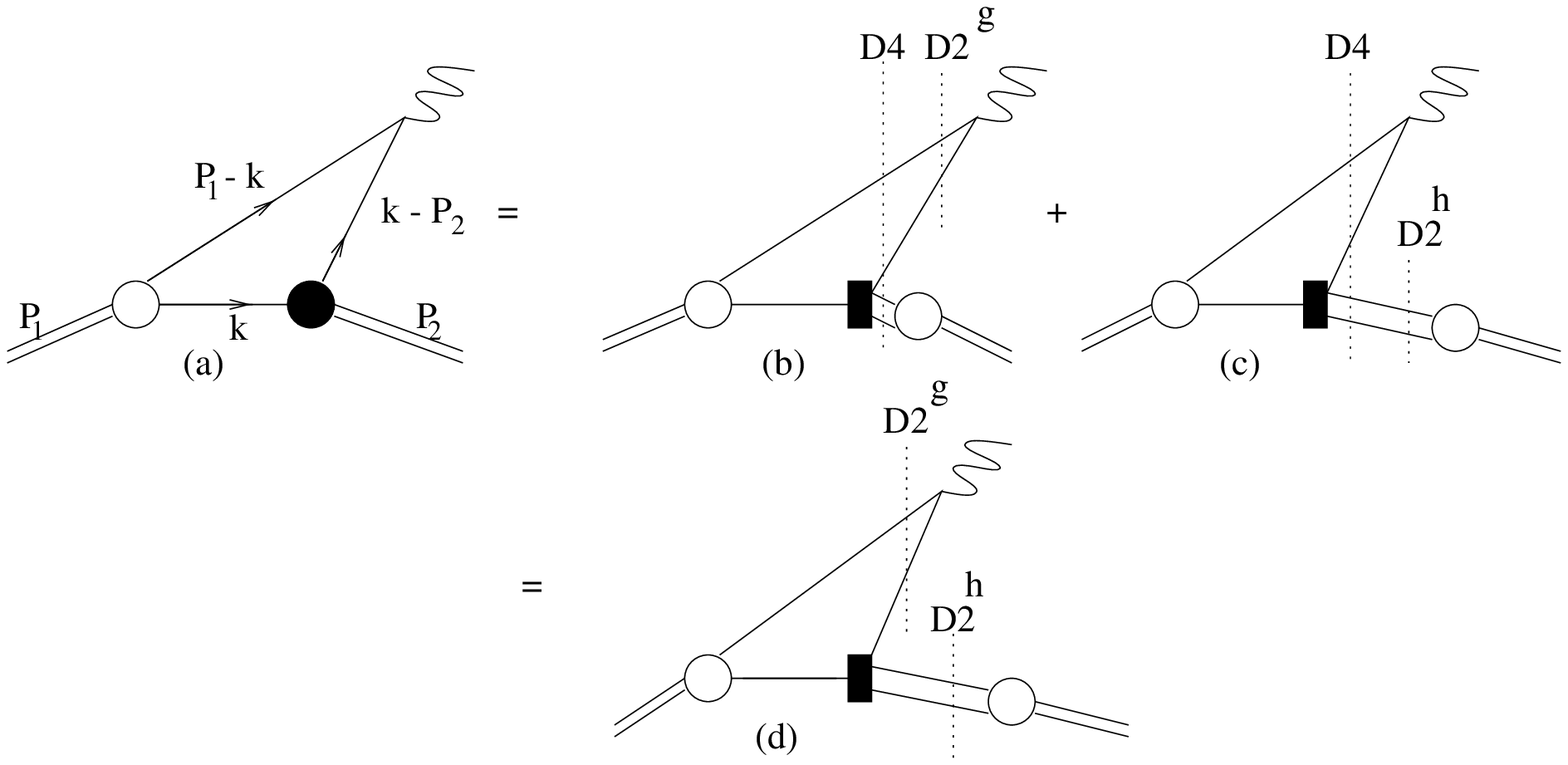,height=2.0in,width=4.in}}
\vspace*{13pt}
\fcaption{Effective treatment of the light-front nonvalence diagram.}
\end{figure}
In this talk, we thus present the way of handling the nonvalence
contribution. Our aim of new treatment is to make the program more
suitable for the CQM phenomenology. As an application of our method, 
we investigate the semileptonic decay processes such as 
$K_{\ell3}$($\ell=e$ or $\mu$) decays since the light-to-light decay
processes bear the largest contribution from the nonvalence part.
Also, the experimental data are best known in theses processes.
Including the nonvalence contribution, our results not only show a
definite improvement in comparison with the experimental data
but also exhibit a covariance (i.e. frame-independence)
of our model.

Thu current matrix element of the semileptonic pseudoscalar to pseudoscalar
meson decays involve the two form factors:
\begin{equation}{\label{eq:K1}}
J^\mu(0)=\langle P_{2}|\bar{Q_{2}}\gamma^{\mu}Q_{1}|P_{1}\rangle
= f_{+}(q^{2})(P_{1}+P_{2})^{\mu} + f_{-}(q^{2})q^{\mu}, 
\end{equation}
where $q^\mu=(P_{1}-P_{2})^\mu$ is the four-momentum transfer to
the lepton pair ($\ell\nu$) and $m^{2}_\ell\leq q^2\leq (M_{1}-M_{2})^{2}$.

To illustrate our method~\cite{JC}, we treat the nonvalence state using the 
Schwinger-Dyson equation to connect the embedded-state shown as the black blob
in Fig.~1(a) to the ordinary LF wave function (white blob in
Fig.~1(d)). To make the program successful, we need some relevant
operator connecting one-body to three-body sector shown as the black box
in Fig.~1(d).
The relevant operator is in general dependent on the involved momenta.
Our {\bf main observation} is that we can remove the four-body energy 
denomenator $D_4$ using the identity $1/D_4D^g_2 + 
1/D_4D^h_2=1/D^g_2D^h_2$ of the energy denominators(see Figs.~1(b) and~1(c))
and obtain the identical amplitude in terms of
ordinary LF wave functions of photon and hadron (white blob) as shown
in Fig.~1(d).
For the small momentum transfer, perhaps the relevant operator
may not have too much dependence on the involved
momenta and one may approximate it as a constant operator.
In contact interaction case, we verified that our prescription of a
constant operator in Fig.~1(d) is an exact solution of
Fig.~1(a).

\begin{table}[htbp]
\tcaption{\label{kpi}
Results for the parameters of $K^{0}_{\ell3}$ decay form factors.}
\centerline{\footnotesize\smalllineskip
\begin{tabular}{|c|c|c|c|c|}\hline
 &\multicolumn{2}{c|}{$q^+\neq0$ frame}& $q^+=0$ frame& \\
\cline{2-4}
 &{\bf Effective(val + nv)} &valence & valence & Experiment~\cite{PDG}\\
\hline
$f_{+}(0)$ & 0.962 & 0.962 & 0.962 & \\
\hline
$\lambda_{+}$& 0.026 & 0.083 & 0.026 & $0.0288\pm0.0015[K^{0}_{e3}]$\\
\hline
$\lambda_{0}$& 0.025 & $-0.017$& 0.001
& $0.025\pm0.006[K^{0}_{\mu3}]$\\
\hline
$\xi_{A}$& $-0.013$ & $-1.10$& $-0.29$
& $-0.11\pm0.09[K^{0}_{\mu3}]$\\
\hline
\end{tabular}}
\end{table}
In Table~\ref{kpi}, we summarize the experimental observables for the
$K_{\ell3}$ decays, where $\lambda_i=M^{2}_{\pi}f'_{i}(0)/f_{i}(0)(i=+,0)$
and $\xi_{A}=f_-(0)/f_+(0)$. We use our linear potential parameters~\cite{CJ} 
in this analysis. As one can see in Table~\ref{kpi}, our new results (column
2) for the slope $\lambda_0$ of $f_0(q^2)=f_+(q^2)+q^2 f_-(q^2)/(M^2_1-M^2_2)$ 
at $q^2=0$ and $\xi_{A}$=$f_-(0)/f_+(0)$
are now much improved and comparable with the data.
Especially, our result of $\lambda_0$= 0.025
obtained from our effective calculation is in excellent
agreement with the data, $\lambda^{\rm Exp.}_0$=0.025$\pm$0.006.
We should note that the form factor $f_+(q^2)$ obtained from $J^+$
in $q^+=0$ frame is immune to the zero-mode contribution 
but the form factor $f_-(q^2)$ obtained from $J_\perp$ receives zero-mode 
contribution, i.e. the differences of $\lambda_0$ and $\xi_A$ between 
our effective solutions and the $q^+=0$ frame results indicate the zero-mode 
contributions. 

In summary, we presented an effective treatment of the LF
nonvalence contributions crucial in the timelike exclusive processes.
Using a SD-type approach and summing the LF time-ordered amplitudes,
we obtained the nonvalence contributions in terms of ordinary LF
wavefunctions of gauge boson and hadron
that have been extensively tested in the spacelike exclusive processes.
Including the nonvalence contribution, our results show a definite
improvement in comparison with experimental data on $K_{\ell3}$
decays. The frame-independence of our
results also indicate that a constant relevant operator is an approximation
appropriate to the small momentum transfer processes. Applications
to the heavy-to-light decay processes involving large
momentum transfers would require an improvement of the relevant operator
including the momentum dependence.
Consideration along this line is underway.

\nonumsection{Acknowledgements}
\noindent
This work was supported by the US DOE under contracts DE-FG02-96ER40947.
The North Carolina Supercomputing Center and the National Energy Research
Scientific Computer Center are also acknowledged for the grant of
supercomputer time.


\end{document}